\documentclass[aps,prb,reprint,superscriptaddress,amssymb,amsmath]{revtex4-1}
\usepackage{graphicx}
\usepackage{dcolumn}
\usepackage{bm, color}
\usepackage[none]{hyphenat} 
\usepackage{bbm}
\usepackage{braket}
\usepackage{tabularx}

\begin{document}
\title{Topological Weyl semimetals in Bi$_{1}$$_{-}$$_{x}$Sb$_{x}$ alloys}
\author{Yu-Hsin Su}
\affiliation{Max Planck Institute for Chemical Physics of Solids, D-01187 Dresden, Germany}
\author{Wujun Shi}
\affiliation{Max Planck Institute for Chemical Physics of Solids, D-01187 Dresden, Germany}
\affiliation{School of Physical Science and Technology, ShanghaiTech University, Shanghai 200031, China}
\author{Claudia Felser}
\affiliation{Max Planck Institute for Chemical Physics of Solids, D-01187 Dresden, Germany}
\author{Yan Sun}
\email{ysun@cpfs.mpg.de}
\affiliation{Max Planck Institute for Chemical Physics of Solids, D-01187 Dresden, Germany}

\begin{abstract}
We have investigated the Weyl semimetal (WSM) phases in bismuth antimony (Bi$_{1-x}$Sb$_{x}$) alloys by the combination of atomic composition and
arrangement. Via first principles calculations, we have found two WSM states
with the Sb concentration of $x=0.5$ and $x=0.83$ with specific inversion
symmetry broken elemental arrangement. The Weyl points are close to the
Fermi level in both of these two WSM states. Therefore, it has a good
opportunity to obtain Weyl points in Bi-Sb alloy. The WSM phase provides a
reasonable explanation for the current transport study of BiSb alloy with
the violation of Ohm's law [Dongwoo Shin \textit{et al}., Nature Materials
16, 1096 (2017)]. This work shows that the topological phases in Bi-Sb
alloys depend on both elemental composition and their specific arrangement.
\end{abstract}

\maketitle

\affiliation{Max Planck Institute for Chemical Physics of Solids, D-01187
Dresden, Germany}

\affiliation{Max Planck Institute for Chemical Physics of Solids, D-01187
Dresden, Germany} 
\affiliation{School of Physical Science and Technology,
ShanghaiTech University, Shanghai 200031, China}

\affiliation{Max Planck Institute for Chemical Physics of Solids, D-01187
Dresden, Germany}

\affiliation{Max Planck Institute for Chemical Physics of Solids, D-01187
Dresden, Germany}

\section{Introduction}

Weyl semimemetal (WSMs) is topological metallic state with valence bands and
conduction bands linearly touching in three dimensional momentum space via
the Weyl points~\cite{Wan2011,volovik2003universe}. These Weyl points are
doubly degeneracy and behave as the monopoles of Berry curvature with
positive and negative charilities, resulting in non-zero topological
charges. Similar to topological insulators (TIs), WSMs also have topological
protected surface states. Owing to the non-zero Berry flux between one pair
of Weyl points with opposite chirality, the surface states present as
non-closed Fermi arc connecting this pair of Weyl points, which have been
observed in several WSMs via Angle-resolved photoemission spectroscopy
(ARPES) and Scanning tunneling microscope (STM) ~\cite{Xu2015TaAs,Lv2015TaAs,Yang2015TaAs,Liu2016NbPTaP,Xu2015NbAs,Belopolski2016NbP,Xu2016TaP,Souma2015NbP,Inoue2016, Batabyal2016, Zheng2016}. Beside Fermi arc surface states, WSMs also host exotic transport
properties in the bulk, such as chiral anomaly effect ~\cite{Huang2015anomaly,Zhang2016ABJ,Wang2015NbP,Niemann2017}, large
magnetoresistance (MR)~\cite{Shekhar2015,Ghimire2015NbAs,Huang2015anomaly,Zhang2016ABJ,Wang2015NbP,Luo2015,Moll2015}%
, strong spin and anomalous Hall effect~\cite{Burkov:2011de,Xu2011,Sun2016,Enk_2017,Shi_2018}, gravitational anomaly
effect~\cite{Gooth2017}, even special catalyst effect~\cite{Rajamathi2017}.

There are already several WSMs were predicted, and some of them were
experimentally verified by the observation of Fermi arcs on the surface ~\cite{Xu2015TaAs,Lv2015TaAs,Yang2015TaAs,Liu2016NbPTaP,Xu2015NbAs,Belopolski2016NbP,Xu2016TaP,Souma2015NbP,Inoue2016, Batabyal2016, Zheng2016}
and negative MR in the bulk transports~\cite{Huang2015anomaly,Zhang2016ABJ,Wang2015NbP,Niemann2017}. So far, most of the
study about inversion symmetry broken WSMs are focused on the space group
without inversion center. Besides the intrinsic space group without
inversion symmetry, alloy is another efficient way to break inversion
symmetry. Owing to the alloy-able of lots of semimetals, it offers a good
opportunity to achieve WSMs in a large number of materials. That motives us
to look back of the well known topological semimetals and semiconductors
based on alloy, in which a typical example is the Z$_{2}$ TI Bi$_{1-x}$Sb$_{x}$~\cite{fu_2007,Hsieh_2008,Hasan:2010ku}.

Bi$_{1-x}$Sb$_{x}$ is the first experimentally discovered 3D Z$_{2}$ TI~\cite{Hsieh_2008}. Due to the similar lattice parameters for Bi and Sb, Bi$_{1-x}$Sb$_{x}$ alloy can be experimentally formed and the composition can be
artificially adjusted. Through atom substitution of Bi by Sb, a non-smooth
band inversion happens along with the topological phase transition at a Sb
concentration of $\approx 0.04$ ~\cite{Hsieh_2008}. The non-trivial band
order can exist in a large range of Sb concentration after the phase
transtion~\cite{Teo_2008}. Since there is not a global band gap for some Sb
concentrations, the Bi$_{1-x}$Sb$_{x}$ presents as a topological semimetal
with inverted band gap~\cite{fu_2007}. In previous studies, the theoretical
studies for Bi$_{1-x}$Sb$_{x}$ were mainly based on the virtual crystal
approximation keeping inversion symmetry as that in Bi and Sb~\cite{Teo_2008, Zhang_BiSb_2009}. However, owing to the chemical difference of
elements Bi and Sb, the alloy of them should break the inversion symmetry.
In combination with the inverted band order and semimetallic feature, the
inversion symmetry broken arrangement in the alloy provides large
possibility to realize WSM.

Recently, the chiral anomaly induced negative MR and violation of Ohm's law
were observed in the transport measurements for Bi-Sb alloy, showing the
emergence of WSM phase~\cite{Shin_2017,Kim_2013}. However, there is still a
lack of fully investigation of the influence of the atomic arrangements in
Bi-Sb alloy on their topological properties. In our present work, a
systematic study on topological nature about the effect of composition and
atomic arrangement in Bi-Sb alloy have been proposed and provides a detailed
classification depending on the sort of topological phase, based on the
quantitative first-principles calculations.

\section{Methods}

The density functional theory (DFT)-based first-principles calculations were
performed by projected augmented wave (PAW) method as implemented in the
Vienna \textit{Ab-initio} Simulation Package (VASP)~\cite{kresse1996}. The
exchange-correlation energy was considered in the generalized gradient
approximation (GGA), following the Perdew-Burke-Ernzerhof (PBE)~\cite{perdew1996} parametrization scheme. The van der Waals interactions were
also taken into account due to the layered lattice structure. In order to
analyze the topological properties, we have projected the Bloch wavefuctions
into maximally localized Wannier functions (MLWFs)~\cite{Mostofi2008}. The
tight-binding model Hamiltonian parameters are determined from the MLWFs
overlap matrix. The surface state was considered under an open boundary
conditions with the half-infinite model using the iterative Green's function
method~\cite{Sancho1984,Sancho1985}.

\section{Results}

\begin{figure}[tbh]
\centering
\includegraphics[width=0.46\textwidth]{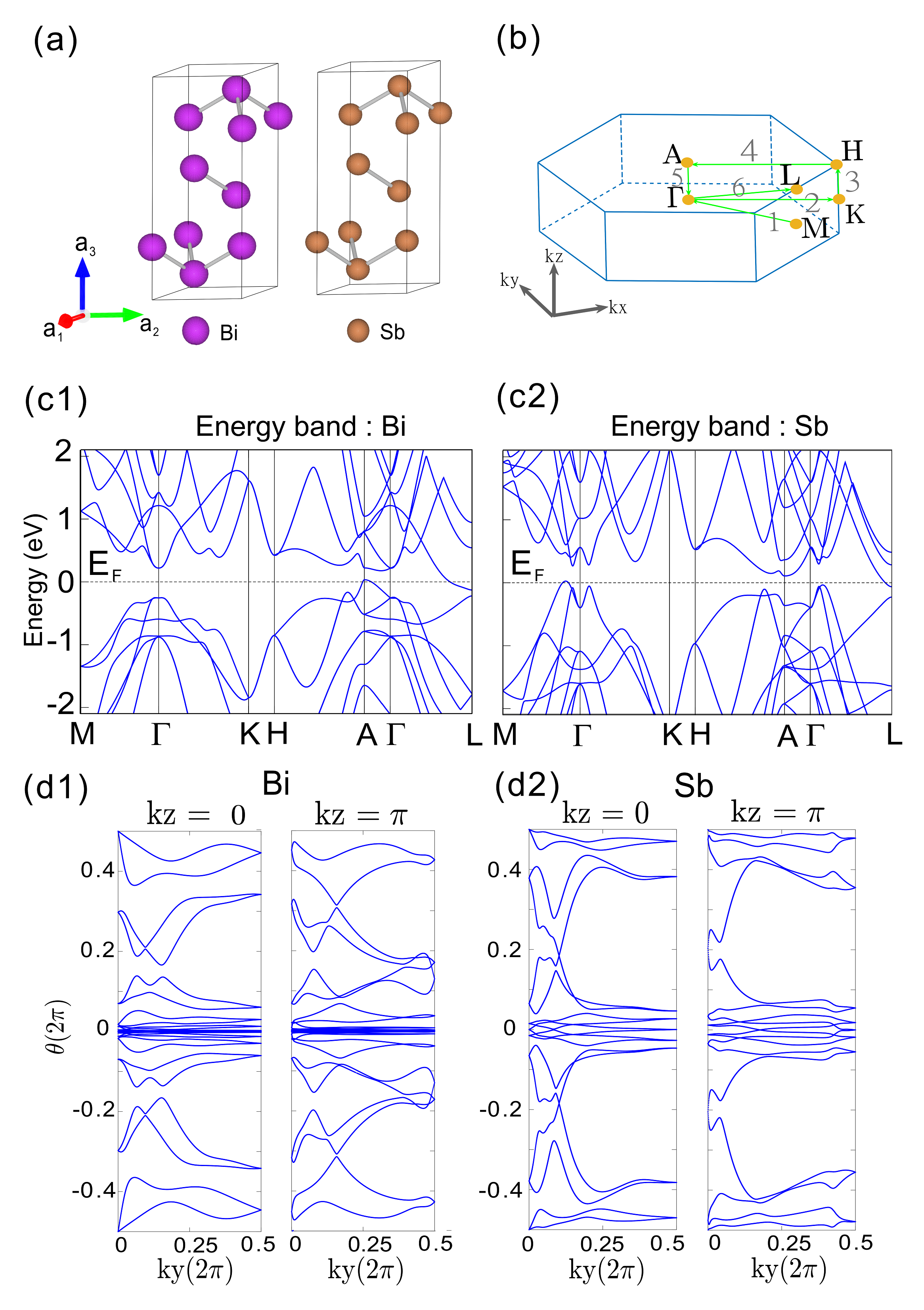}
\caption{ (a)The hexagonal conventional unit cell of Bi and Sb atoms,
consists of layered structure. The grey solid lines denote the intralayer
bonding. (b) The Brillouin zone of hexagonal unit cell and the k-points
path, containing high symmetry points, along which the band structure is
calculated. (c) and (d) are the corresponding bulk energy band. (e, f) The
evaluated Wannier centers on the plane k$_{\text{z }}$= 0 and k$_{\text{z }}$= $\protect\pi $, respectively. }
\label{fig:elment}
\end{figure}

Both Bi and Sb have the rhombohedral A7 crystal structure with space group $R\overline{3}m$, as indicated in Fig. 1(a). Consistent with previous reports,
Bi and Sb have similar electronic band structures due to their similar
chemical properties, see Fig. 1(c-d). Whereas, owing to different strengh of
spin-orbital coupling (SOC), they have different band orders and therefore
different Z$_{2}$ topological invariant of $\nu _{0}$ = 0 and 1,
respectively ~\cite{Teo_2008, Zhang_BiSb_2009}, see Fig. 1(e-f). Based on
the hexagonal lattice, we have investigated all atomic arrangements for
different elements composition of x = 0.17, 0.33, 0.5, 0.67, and 0.83. In
our calculations the band inversion between antisymmetric ($L_{a}$) and
symmetric ($L_{s}$) orbitals happens at the concentration with x = 0.5,
which agrees well with previous tight binding analysis. Above that the
system present as topological insulators or semimetals, in which the WSM
states were found in one arrangement for the composition of x = 0.5 and x =
0.83, respectively. For convenience, we will take the WSM phase in the
compositions of x = 0.5 as the example to introduce the electronic structure.

\begin{figure}[tbh]
\centering
\includegraphics[width=0.46\textwidth]{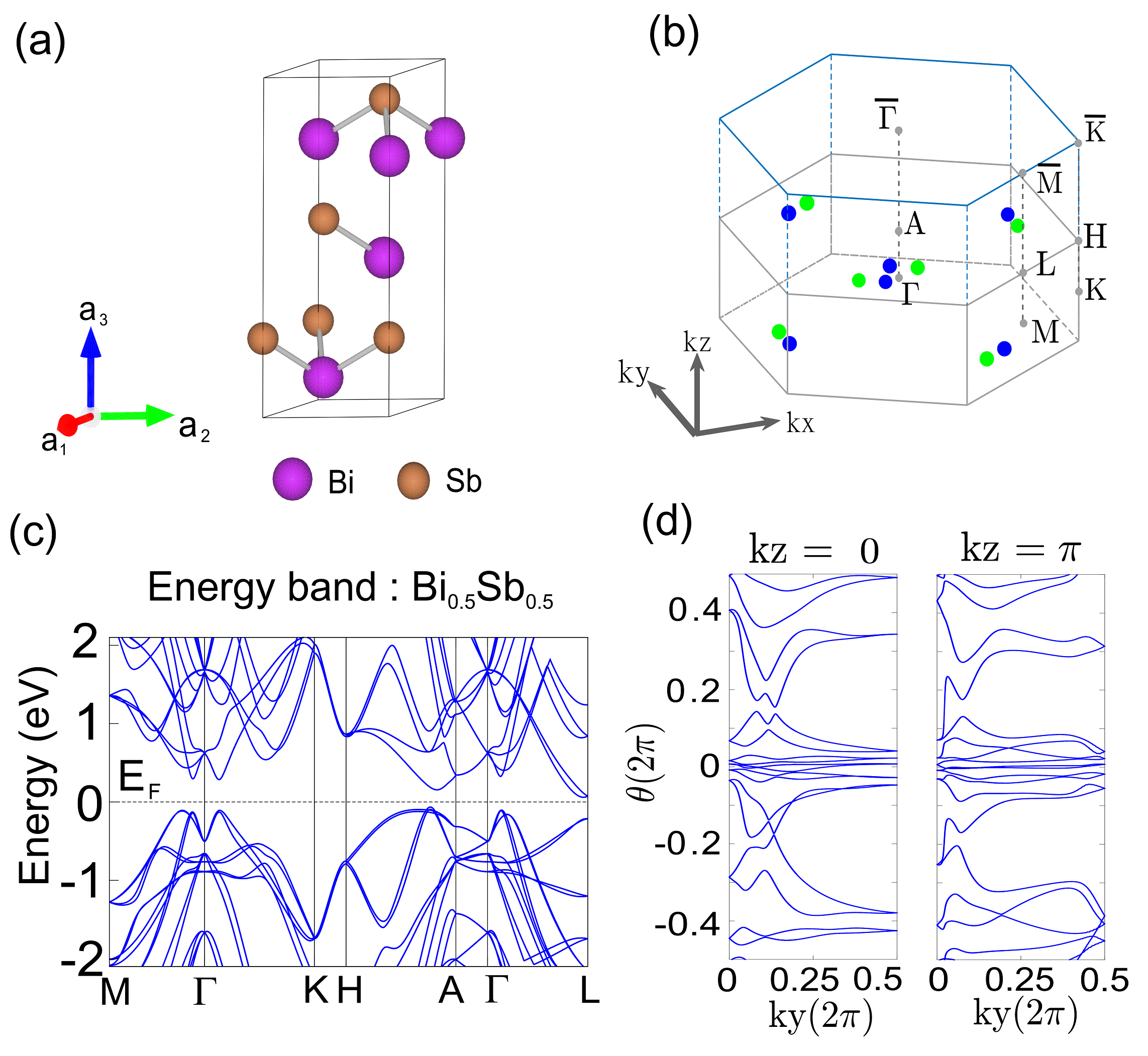}
\caption{ (a) The fully relaxed crystal structure for Bi$_{0.5}$Sb$_{0.5}$
with the specific arrangement of Bi and Sb atoms. (b) The 3D BZ of hexagonal
unit cell where the location of Weyl points with positive chirality (blue)
and negative chirality (greeen) are shown, and projected 2D BZ of (001)
surface. (c) Bulk Energy band for specific Bi$_{0.5}$Sb$_{0.5}$ . (d) The
evaluated Wannier centers on the plane k$_{\text{z}}$ = 0 and k$_{\text{z}}$
= $\protect\pi $, respectively. }
\label{fig:half}
\end{figure}

Fig. 2(a) shows the lattice structure of Bi$_{0.5}$Sb$_{0.5}$ with the
specific atomic arrangement, where Bi layer and Sb layer stacking with each
other in c direction. Though the formation energy is slightly higher ($\sim
3 $ meV per formula unit) than that in the ground state arrangement, the
energy difference is almost the limit of accuracy of the DFT itself.
Therefore, it is reasonable to obtain this atomic arrangement in
experiments. Owing to the absence of inversion symmetry, the spin degeneracy
splits for all the bands, see Fig. 2(c), which also provides the possibility
of Weyl points. The electronic band structure along high symmetry lines
exhibits as a direct band gap insulator. To check the topological phase, we
calculated the Wannier center evolutions in the high symmetry planes of k$_{%
\text{z}}=0$ and $\pi $, respectively. As shown in Fig. 2(d), the Wannier
center curves calculated on the plane k$_{\text{z}}$ $=0$ presents the
existence of discontinuity, yet on plane k$_{\text{z}}$ $=\pi $ plane, the
curves are always continuously connected. It indeed provides the evidence of
a topological non-trivial phase in Bi$_{0.5}$Sb$_{0.5}$.

\begin{figure}[tbh]
\centering
\includegraphics[width=0.46\textwidth]{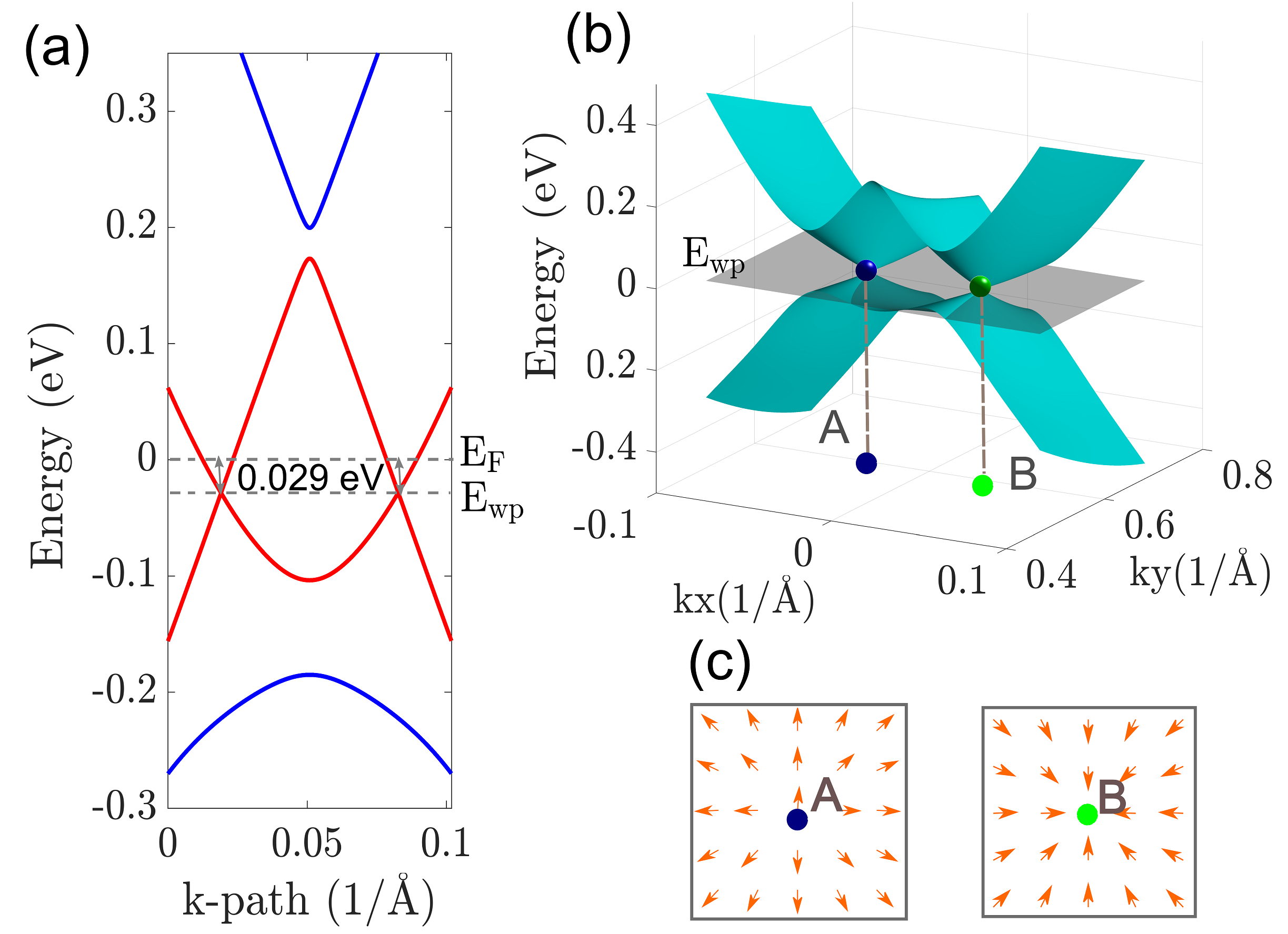}
\caption{ (a) The energy band passing through two Weyl points, where E$_{%
\text{F}}$ and E$_{\text{wp}}$ represent the Fermi energy and the energy of
Weyl points, respectively. The energy bands with color red denote the
highest occupied band and the lowest unoccupied band. (b) The energy band
structure around two Weyl points A and B on the plane k$_{\text{z}}$=0.4620(1/\AA ). (c) The Berry curvature calculated on the two Weyl points A
and B with normalization. }
\label{fig:WP}
\end{figure}

Though there is a general gap along high symmetry lines, its density of
states (DOS) is not zero at Fermi level, implying a topological semimetal
with bands cutting the Fermi level at some lower symmetry directions.
Indeed, we found one pair of linear band crossings away from high symmetry
points at (-0.031822, 0.592594, 0.245628) ($1/${\AA }) and (0.031808,
0.592380, 0.245608) ($1/${\AA }) as shown in Fig. 3. This pair of linear
crossing points are Weyl cones behaving as the sink and source of Berry
curvature. Considering the C3 rotation and time reversal symmetry, there are
6 pairs of Weyl points in total. Since the Weyl points is only around 30 meV
bellow Fermi level, the Weyl points dominated phenomenons should be easy to
detect by both surface technique and bulk transport measurements.

\begin{figure}[tbh]
\centering
\includegraphics[width=0.46\textwidth]{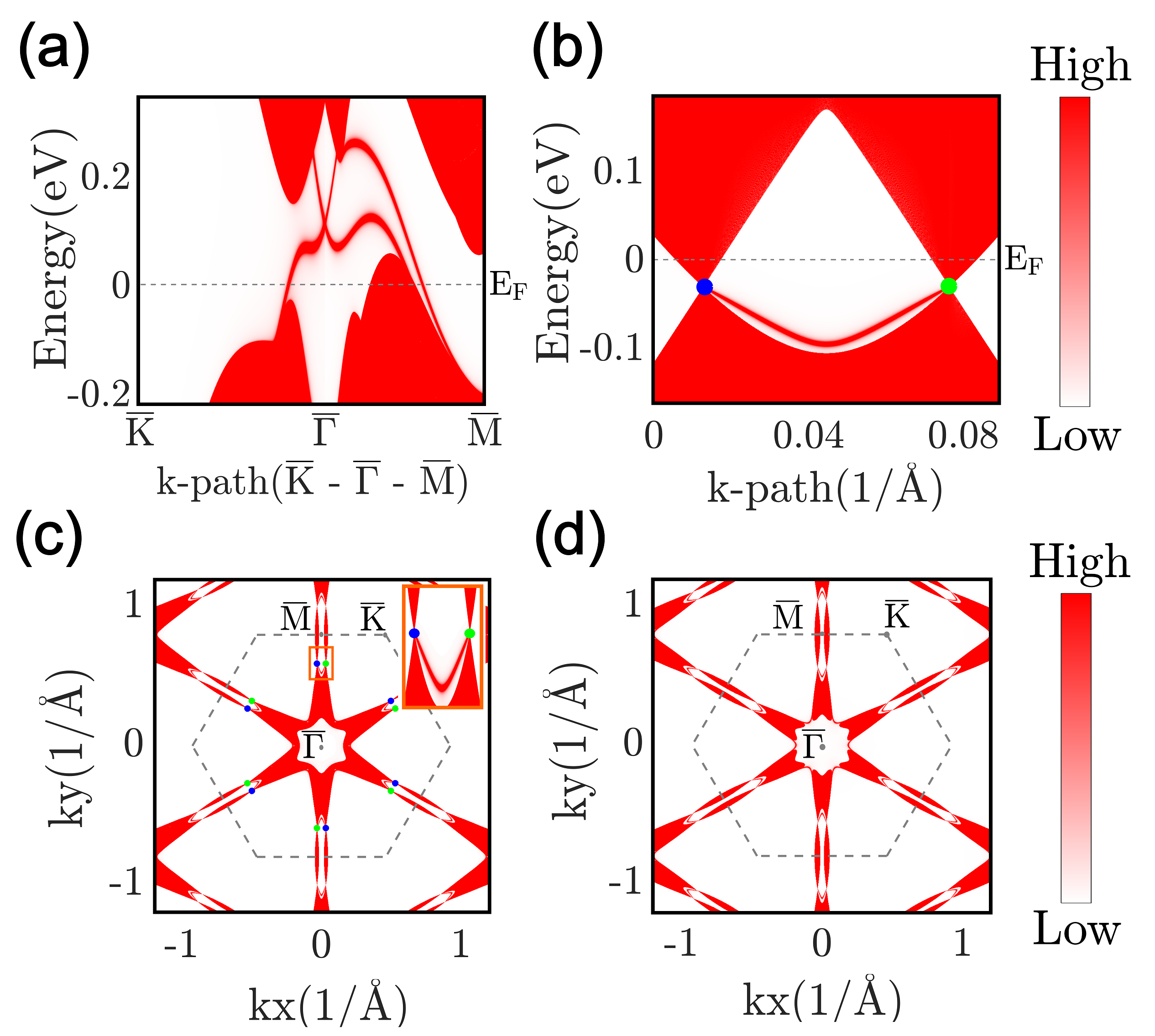}
\caption{ (a) The local DOS of (001) plane calculated along a specific path
passing through high symmetry ponits ($\overline{\text{K}}$-$\overline{\Gamma }$
-$\overline{\text{M}}$), where E$_{\text{F}}$ denotes the Fermi enetrgy.
(b) The local DOS of (001) plane calculated along a pth passing through two
Weyl points. (c) The surface DOS on the (001) plane at the energy of Weyl
points E$_{\text{wp}}$, where the Weyl points with positive (negative)
chirality are denoted as blue (green) circles. and the grey dashed line
represents the first BZ of the surface. The inset is the enlarged region
around a pair of Weyl points. (d) The surface DOS on the plane (001) at the
Fermi energy. }
\label{fig:arc}
\end{figure}

An important characteristic of WSM is the presence of surface Fermi arc
states~\cite{Wan2011}. Considering the easy cleave-able bonding, we have
chosen the (001) surface in our analysis. Fig. 4(a) and (b) illustrate the
local DOS along two particular paths, one is passing through high symmetry
lines of $\overline{\text{K}}$-$\overline{\Gamma }$-$\overline{\text{M}}$,
and the other is across a pair of Weyl points with opposite chirality. Along
the high symmetry lines of $\overline{\text{K}}$-$\overline{\Gamma }$-$\overline{\text{M}}$ one can see that, besides the surface Dirac cone at $\overline{\Gamma }$ points, there is the other surface bands on $\overline{\Gamma }$-$\overline{\text{M}}$ line connecting bulk conduction and valence
states. Since one pair of Weyl points with opposite chirality symmetrically
locates on the two sides of $\overline{\Gamma }$-$\overline{\text{M}}$ line,
the additional surface bands should be the Fermi arc related states. To
check this speculation, we directly plot surface energy dispersion along the
k-path crossing one pair of Weyl points and perpendicular to $\overline{\Gamma }$-$\overline{\text{M}}$. As shown in Fig. 4(b), a surface bands
merges into bulk via the Weyl points, just the typical feature for the Fermi
arc related states. Fixing the energy at Weyl points, one can easily found
the Fermi arc starting from one Weyl points and end at the other see Fig.
4(c). Moreover, the Weyl points present as the linear touching of the
projected bulk states. Hence, the Bi$_{0.5}$Sb$_{0.5}$ is a type-II WSM.
From the energy dispersions along $\overline{\Gamma }$-$\overline{\text{M}}$
and one pair of Weyl points in Fig. 4(a-b), the Fermi arc related states can
range in a large energy window from ~$-$0.1 to 0.2 eV, hence the Fermi arcs
can be also observed at Fermi level, see Fig. 4(e). Therefore, the WSMs in Bi$_{0.5}$Sb$_{0.5}$ is further confirmed by the surface Fermi arcs.

Besides Bi$_{0.5}$Sb$_{0.5}$, the composition $x=0.87$ also shows WSMs state
with Weyl points lying at ($-$0.00436, 0.721258, 0.275386) ($1/${\AA })
and 0.1 eV bellow Fermi energy. Based on the hexagonal cell, in total, there
are five different compositions by atom substitutions for the alloy, and two
of them can host WSM phase. Therefore, it has a large possibility to archive
Weyl points close to Fermi level in Bi-Sb alloy, and our DFT calculations is
in good agreement with the unusual transport properties observed in
experiments~\cite{Shin_2017,Kim_2013}. 

\section{\textrm{Conclusion}}

\textrm{In conclusion, we investigated the influence of atomic composition
and arrangement on the topology of Bi$_{1-x}$Sb$_{x}$ alloy via $ab-initio$
calculations. Increasing the concentration of Sb, topological phase
transition happens in Bi$_{1-x}$Sb$_{x}$. Interestingly, the atomic
arrangements for a particular compositions in Bi$_{1-x}$Sb$_{x}$ remarkably
relates to its own topological feature in contrast to the previous
researches that just emphasizes the effect of composition. As a result, two
WSM states were obtained at $x=0.5$ and $x=0.87$ with specific atomic
arrangement with the absence of inversion symmetry. Our result is helpful
for the understanding of recently unusual transport properties observed in
experiments. This work reveals the importance of the combination of
elemental composition and their specific arrangement for the comprehensive
understanding of topological phases in Bi-Sb alloys. }

\begin{acknowledgments}
This work was financially supported by the European Research Council (ERC) Advanced Grant
(No. 291472) 'IDEA Heusler' and ERC Advanced Grant (No. 742068) 'TOPMAT'.
\end{acknowledgments}

\bibliography{topmater.bib}

\end{document}